\documentstyle[emulateapj,times,epsfig]{article}

\newcommand{\pfrac}[2]{\left(\frac{#1}{#2}\right)}

\def\Mesz{M\'esz\'aros~}
\def\Pacz{Paczy\'nski~}

\def\beq{\begin{equation}}
\def\enq{\end{equation}}
\def\bea{\begin{eqnarray}}
\def\ena{\end{eqnarray}}
\def\bec{\begin{center}}
\def\enc{\end{center}}
\def\etal{{et al.~}}
\def\eps{\epsilon}

\def\pair{$e^{\pm}$}
\def\kp{k_\pm}
\def\k2{k_{\pm,2}}
\def\Npm{N_\pm}

\def\cm2si{\hbox{cm$^{-2}$s$^{-1}$}}

\begin{document}

\title{Pair loading in Gamma-Ray Burst Fireball And Prompt Emission From Pair-Rich Reverse Shock}
\author{Zhuo Li$^{1,2}$, Z. G. Dai$^1$, T. Lu$^1$ and L. M. Song$^2$}
\affil{ $^1$Department of Astronomy, Nanjing University, Nanjing
210093, China\\
$^2$Particle Astrophysics Lab., Institute of High Energy Physics,
Chinese Academy of Sciences, Beijing 100039, China}

\begin{abstract}
Gamma-ray bursts (GRBs) are believed to originate from
ultra-relativistic winds/fireballs to avoid the ``compactness
problem". However, the most energetic photons in GRBs may still
suffer from $\gamma-\gamma$ absorption leading to \pair~ pair
production in the winds/fireballs. We show here that in a wide
range of model parameters, the resulting pairs may dominate those
electrons associated with baryons. Later on, the pairs would be
carried into a reverse shock so that a shocked pair-rich fireball
may produce a strong flash at lower frequencies, i.e. in the IR
band, in contrast with optical/UV emission from a pair-poor
fireball. The IR emission would show a 5/2 spectral index due to
strong self-absorption. Rapid responses to GRB triggers in the IR
band would detect such strong flashes. The future detections of
many IR flashes will infer that the rarity of prompt optical/UV
emissions is in fact due to dust obscuration in the star formation
regions.
\end{abstract}


%%%%%%%%%%%%%%%%%%%%%%%%%%%%%%%%%%%%%%%%%%%%%%%%%%%%%%%%%%%%%%%%%%%%%%%%%%%%%%%%%%%%%
\section{Introduction}
There is little doubt that GRBs arise from relativistically
expanding winds/fireballs, which also drive relativistic blast
waves that responsible for later time emission at X-ray, optical
and radio bands (see reviews of Piran 1999; van Paradijs,
Kouveliotou \& Wijers, 2000; Cheng \& Lu 2001; and \Mesz 2002).
The observed GRB spectra are hard, with a significant fraction of
the energy above the \pair~ pair formation energy threshold, so it
is believable that there are processes of substantial pair
production related to GRBs. One case of \pair~ pair production in
GRBs is that the GRB photons are back-scattered by the ambient
medium and interact with other outgoing high-energy GRB photons,
leading to \pair~ pair production and then deposition of momentum
into the external medium. Such a case has been considered by Madau
\& Thompson (2000); Thompson \& Madau (2000); Dermer \& B\"ottcher
(2000); Madau, Blandford \& Rees (2000); \Mesz, Ramirez-Ruiz \&
Rees (2001); Beloborodov (2002); Ramirez-Ruiz, MacFadyen \&
Lazzati (2002), which could be called an ``external case''. In
this paper, we consider the so-called ``internal case'', in which
pairs are produced within the relativistic fireball due to
$\gamma-\gamma$ absorption between high energy photons, and the
formed pairs remain mixing in the fireball matter. The pairs in
the external medium (external case) will modify the usual
properties of the forward shock and then the later afterglow,
while the pairs in the wind/fireball (internal case) affect the
reverse shock that is responsible for prompt emission in soft
bands during the bursting phase. Pilla \& Loeb (1998) have also
studied spectral signatures of GRBs by considering pair production
in fireballs. It is possible that the pairs are so abundantly
generated that the fireball becomes optically thick, and the GRB
photons undergo Compton multi-scattering before leaking out. \Mesz
\etal (2002) discussed this case to account for the currently
known X-ray flashes.

A large Lorentz factor is required to avoid the ``compactness
problem'' in GRBs (see, e.g., Piran 1999). However, the
high-energy photons may still suffer from absorption, and the
generated pairs may dominate the electrons associated with
baryons, and then change the conditions in the reverse shock. The
presence of a large number of $e^{\pm}$ pairs decreases the shared
energy per lepton in the reverse shock, so that the reverse flash
is softened to lower frequencies. The reverse-shock model (Sari \&
Piran 1999; \Mesz \& Rees 1999) has been successful to interpret
the strong optical flash detected in GRB 990123 (Akerlof \etal
1999). Here we show that a pair-rich reverse shock gives rise to
even stronger radiation in longer wavelength, e.g. the IR, as
opposed to the UV/optical. We discuss in \S 2 the pair production,
and in \S 3 the effect of pairs on the prompt flashes from reverse
shocks. We discuss the current observation of prompt optical
radiation in \S 4. Conclusions and discussion on future
observations are given in \S 5.

%%%%%%%%%%%%%%%%%%%%%%%%%%%%%%%%%%%%%%%%%%%%%%%%%%%%%%%%%%%%%%%%%%%%%%%
\section{Pair loading in GRBs}

%%%%%%%%%%%%%%%%%%%%%%%%%%%%%%%%%%%%%%%%%%%%%%%%%%%%%%%%%%%%%%%%%%%%%%%%%
\subsection{GRB emission site}
Consider a GRB central engine producing a relativistic wind
outflow, which carries energy from the center to some radius that
GRB emission arises. We assume the isotropic geometry, which is
valid even for a jetted GRB as long as the Lorentz factor
$\Gamma>1/\theta_{jet}$, with $\theta_{jet}$ the jet opening
angle. The width of outflow is $\Delta\la 10^{12}$~cm for a wind
lasting a duration of $\la100$~s. The emission energy may come
from two energy forms, the kinetic energy of the baryon bulk and
the magnetic energy of the fireball. Therefore the initial
wind/fireball energy $E$ may be divided into two components (e.g.
Zhang \& \Mesz 2002). The first component is the bulk kinetic
energy $E_K$ that, as considered in the simplest fireball model,
comes from the thermal energy of photons, pairs and baryons after
initial acceleration. The second component is the energy $E_P$ of
the Poynting flux from the central engine. Define $\sigma\equiv
E_P/E_K$, then the fireball is Poynting-flux dominated if
$\sigma\ga 1$, as suggested by recent detection of extreme
polarization in GRB 021206 (Coburn \& Boggs 2003), or
kinetic-energy dominated if $\sigma\la 1$.

The non-thermal spectra of GRBs require that the site of GRB
emission should be beyond the photospheric radius at which the
outflow becomes optically thin to scattering by electrons
associated with baryons. The total baryonic electron number in the
fireball is $N_b=E/(1+\sigma)\Gamma m_pc^2$, with $\Gamma$ the
Lorentz factor of the fireball. Thus the related optical depth
$\tau_b=\sigma_TN_e/4\pi R^2\la1$ leads to a lower limit to the
emission radius,
\begin{equation}
R_b=3.4\times10^{13}E_{52}^{1/2}\Gamma_{300}^{-1/2}(1+\sigma)^{-1/2}{\rm
cm}.
\end{equation}
Hereinafter, we use c.g.s. units and the convention $U_x=U/10^x$,
except for the Lorentz factors which are scaled as
$\gamma_x=\gamma/x$. Since $R_b\propto\tau_b^{-1/2}$ is
insensitive to $\tau_b$, $R_b$ in the above equation do not change
much even though the GRB photons undergo a few times scattering,
$\tau_b\sim$~a few (not possible to be quite large). It is seen
that usually $R_b\gg\Delta$, so the emission region of GRBs can be
regarded as thin shells.

GRB light curves at gamma-ray energy are often highly variable
with time scale $\delta t\sim1$~ms~$-1$~s. Because of radiation
beaming an observer can see only the part of the emitting fireball
sphere up to an angle $\Gamma^{-1}$ from the line of sight. The
sphere curvature leads to a delay in the photon arrival time by
$R/2\Gamma^2c$. This so-called angular spreading time should not
be larger than the observed variability time scale $\delta t$,
which yields an upper limit to $R$
\begin{equation}\label{Rvar}
R_{var}=5.4\times10^{14}\Gamma_{300}^2\delta t_{-1}{\rm ~cm}.
\end{equation}
Thus, from the properties of GRB spectra and light curves we can
constrain the emission site to the regime
\begin{equation}\label{thinsite}
R_b\la R\leq R_{var}.
\end{equation}
We may take the typical radius as $R=10^{14}R_{14}$~cm in the
following.

%%%%%%%%%%%%%%%%%%%%%%%%%%%%%%%%%%%%%%%%%%%%%%%%%%%%%%%%%%%%%%%%%%%%%%%%%%%%%%%%%%%%%%
\subsection{Pair loading}
For an initially thermal-energy-dominated GRB fireball, pair
production in GRBs begins in the fireball acceleration phase when
the fireball is still opaque to photons, but the pair number
density decreases exponentially with increasing radius during this
phase (Shemi \& Piran 1990). When the fireball becomes transparent
to Compton scattering, the pair number has been negligible
compared with baryonic electrons. We neglect the pair production
in this case.

The intense pair production may occur in the bursting phase due to
$\gamma-\gamma$ absorption between GRB photons. For the most
energetic photons at the high-energy end of the spectrum, the
optical depth of $\gamma-\gamma$ absorption  may exceed unity.
Thus, they will not escape freely from the fireball, and will
produce \pair pairs. The pairs will remain in the fireball, with
the same bulk Lorentz factor as the fireball (static in the
comoving frame). We here calculate how many \pair pairs generated
in this process.

The observed photon spectra of GRBs can be approximated by a
broken power-law, with a high-energy portion of the form
$N_\eps\propto\eps^{-\beta}$ for $\eps>\eps_b$, where
$\eps=h\nu/m_ec^2$ is the photon energy in units of the electron's
rest energy, and $\eps_b\sim 1$ is the break energy above which
the index $\beta\sim2-3$. The photon number above a certain energy
$\eps$ is, therefore, approximated as
\begin{equation}
N_{>\eps}=\pfrac{\beta-2}{\beta-1}\frac{E_\gamma}{\eps_bm_ec^2}\pfrac{\eps}{\eps_b}^{-(\beta-1)},
\end{equation}
where $E_\gamma$ is the GRB energy emitted in gamma-rays.

A photon with energy $\eps$ may annihilate any photons above
$\eps_{an}=\Gamma^2/\eps$. Its optical depth due to
$\gamma-\gamma$ absorption is approximated in an analytical
approach by Lithwick \& Sari (2001) as
\begin{equation}
\tau_{\gamma\gamma}(\eps)=\frac{(11/180)\sigma_TN_{>\eps_{an}}}{4\pi
R^2}.
\end{equation}
A photon $\eps$ with $\tau_{\gamma\gamma}(\eps)>1$ would be
absorbed in the fireball. Setting
$\tau_{\gamma\gamma}(\eps_{cut})=1$ results in a cut-off energy
\begin{equation}\label{cut}
\eps_{cut}=400E_{\gamma,52}^{-{1\over\beta-1}}\eps_b^{-{\beta-2\over\beta-1}}\\
\Gamma_{300}^2R_{14}^{2\over\beta-1},
\end{equation}
where the numerical coefficient on the right-hand side corresponds
to typical parameter $\beta=2.2$ (Preece \etal 2000). The observed
GRB spectrum above cut-off energy $\eps_{cut}$ should be
attenuated. {\em EGRET} on board the {\em Compton-GRO} satellite
have detected several GRBs emitting very high energy gamma-ray
tails. If the cut-off energy can be detected by future satellites
such as {\em GLAST}, we can constrain the emission radius and the
initial Lorentz factor, which are the most important physical
parameters in GRBs. For examples, if no significant attenuation
above 1 GeV, $\eps_{cut}m_ec^2>1$~GeV, this yields a limit
\begin{equation}
R>R_{cut}(1{\rm
~GeV})=2.6\times10^{14}E_{\gamma,52}^{1/2}\eps_b^{(\beta-2)/2}\Gamma_{300}^{-(\beta-1)}{\rm
cm}.
\end{equation}
%All the quantities except for $\eta$ on the right-hand side of Eq.
%(\ref{cut})?? can be determined directly by observations.

The photons above $\eps_{cut}$ would be absorbed for $e^\pm$
production, and each absorbed photon corresponds to a pair of
$e^\pm$. Integrating the photon spectrum above $\eps_{cut}$ we
have the total number of the resulting pairs
\begin{equation}\label{number1}
N_\pm =1.5\times10^{54}
E_{\gamma,52}^2\eps_b^{2(\beta-2)}\Gamma_{300}^{-2(\beta-1)}R_{14}^{-2},
\end{equation}
and hence the ratio
\begin{equation}\label{number}
\kp\equiv N_\pm/N_b=69
E_{\gamma,52}^2\eps_b^{2(\beta-2)}\Gamma_{300}^{-(2\beta-3)}E_{52}^{-1}(1+\sigma)R_{14}^{-2},
\end{equation}
where we have scaled both $E$ and $E_\gamma$ in units of
$10^{52}$~ergs assuming that the efficiency of a GRB is
significant. We note that the calculation of $e^\pm$ number in Eq.
(\ref{number1}) does not depend on detailed models of energy
dissipation in emission region, e.g. internal shock model or
magnetic dissipation model. Our calculation is based on
observations of GRB properties, i.e. spectral and temporal
profiles.

Eq. (\ref{number}) infers that $\kp$ is sensitive to $R$. The
condition for strong pair loading, i.e. $2\kp>1$, is
\begin{equation}\label{Rload}
R<R_{load}=1.2\times10^{15}E_{\gamma,52}\eps_b^{\beta-2}\Gamma_{300}^{-(\beta-{3\over2})}E_{52}^{-1/2}(1+\sigma)^{1/2}{\rm
cm}.
\end{equation}
The case $2\kp>m_p/m_e$ in which pair mass dominates baryon mass
is even achieved if $R<R_{dom}=R_{load}(m_p/m_e)^{-1/2}$. From
Eqs. (\ref{number}) and (\ref{Rload}) we can see that a
wind/fireball with typical parameters generally becomes pair rich
in the bursting phase. These secondary pairs are possible to form
an optically thick screen again for the GRB emissions (e.g.
Guetta, Spada \& Waxman 2001; Kobayashi, Ryde \& MacFadyen 2002;
\Mesz \etal 2002). The optical depth due to Compton scattering by
pairs is $\tau_\pm=\sigma_T(2\Npm)/4\pi R^2$. Substituting Eq.
(\ref{number1}) into $\tau_\pm=1$ leads to
\begin{equation}\label{Rpair}
R_\pm=2.4\times10^{14}E_{\gamma,52}^{1/2}\eps_b^{(\beta-2)/2}\Gamma_{300}^{-(\beta-1)/2}{\rm
cm}.
\end{equation}
A classic GRB takes place in the site optically thin to
scattering, so the emission radius should be large with $R\ga
R_\pm$. Also the permitted regime of Eq. (\ref{thinsite}) for
optically thin GRBs is changed to $R_\pm\la R\leq R_{var}$. If
$R\la R_\pm$, the wind/fireball is optically thick. However the
generated pairs cannot build up an optical depth substantially
exceeding unity, and Compton multi-scattering degrades the photons
into X-ray energy. This gives an interpretation of X-ray flashes
(\Mesz \etal 2002)\footnote{Strictly speaking, the way we
calculate the pair number is only valid for optically thin bursts.
However, $\tau_\pm$ is not large, and the spectral break energy is
degraded no more than an order of magnitude. Furthermore, the
calculation of pair numbers is insensitive to the break energy
provided $\beta\sim2$. Therefore, our calculation for optically
thick case here is still valid.}.

Fig. 1 shows the above special radii in the $\Gamma-R$ diagrams,
where the GRB parameter space is separated into different regions.

%%%%%%%%%%%%%%%%%%%%%%%%%%%%%%%%%%%%%%%%%%%%%%%%%%%%%%%%%%%%%%%%%%%%%%%%%%%%%%%%%%%%%%%%%%
\section{Prompt flashes from pair-rich reverse shocks}

After the bursting phase, the fireball cools into a cold shell,
which continues to expand at an ultrarelativistic speed and with
kinetic energy $E_a$. This afterglow energy should be less than
the initial wind/fireball energy after the burst, and $E_a\la
E-E_\gamma$. When the cold shell runs into the surrounding medium,
the kinetic energy is dissipated into internal energy, which gives
rise to prompt emission and long-term afterglow. Sari \& Piran
(1999a, 1999b) and \Mesz \& Rees (1999) (see also \Mesz, Rees \&
Papathanassiou 1994; \Mesz \& Rees 1997; Panaitescu \& \Mesz 1998)
have discussed the prompt emission in the context of GRB 990123,
assuming no-pair fireball. We discuss here the prompt emission
from a pair-rich fireball.

The interaction between the shell and the surrounding medium forms
two shocks: a reverse shock that propagates into the cold shell,
increasing its internal energy then giving rise to a prompt flash;
and a forward shock that propagates into the cold medium and
produces a long-term multiwavelength afterglow. The shocked medium
and shell material are in pressure balance and separated by a
contact discontinuity. There are four regions of distinct
properties: the unshocked surrounding medium (denoted ``1"), the
shocked medium (``2"), the shocked shell material (``3") and the
unshocked shell (``4").

Now the reverse shock may propagate into a pair-rich shell, with
lepton number density $n_{e,4}=(1+2\kp)n_{b,4}$ and mass density
$\rho_4=n_{b,4}m_p[1+2\kp(m_e/m_p)]$ (both in rest frame of region
4). The shocked material in region 2 and 3 moves together, thus
they have the same Lorentz factor, $\gamma_2=\gamma_3$. The
Lorentz factor of the reverse shock relative to region 4 is
$\gamma_{3,4}\simeq\gamma_4/\gamma_3$. The shock jump conditions
yield that the internal energy and lepton number density of the
shocked shell material are $U_3\simeq4\gamma_{3,4}^2\rho_4 c^2$
and $n_{e,3}=4\gamma_{3,4}n_{e,4}$, respectively (Blandford \&
McKee 1976). Let a fraction $\xi_e$ of the internal energy goes
into leptons, including the baryonic electrons and secondary
pairs. These shocked leptons are expected to be in a power-law
distribution: $dn_{e,3}/d\gamma_e\propto\gamma_e^{-p}$ for
$\gamma_e>\gamma_m$, with $p\approx2$.

The emission reaches a peak when the reverse shock crosses the
inner edge of the shell at
\begin{equation}
t_{cr}=\max[\Delta/c,t_{dec}]
\end{equation}
(Sari \& Piran 1999b) with the deceleration time
\begin{equation}
t_{dec}=3.4E_{a,52}^{1/3}n_1^{-1/3}\gamma_{4,300}^{-8/3}{\rm ~s}.
\end{equation}
The Lorentz factor of shocked shell at $t_{cr}$ could be estimated
as
\begin{equation}\label{gamma3}
\gamma_3=\pfrac{3E_a}{32\pi
n_1m_pc^5t_{cr}^3}^{1/8}=230\pfrac{E_{a,52}}{n_1}^{1/8}t_{cr,1}^{-3/8}.
\end{equation}
%while the relative one $\gamma_{3,4}$ is of the order of unity at $t_{cr}$.
We will focus now on the emission at the peak time
$t_{cr}$.

The characteristic Lorentz factor (in the rest frame of region 3)
of the accelerated leptons is therefore
\begin{equation}\label{gm}
\gamma_m=\xi_e\frac{U_3}{n_{e,3}
m_ec^2}=\xi_e\mu(\gamma_4/\gamma_3),
\end{equation}
where $\mu$ is the effective mass per lepton in units of electron
mass and is given by
\begin{eqnarray}
\frac{\mu m_e}{m_p} & =       & { 1+2\kp(m_e/m_p) \over 1+2\kp}\nonumber\\
        & \simeq & \cases{ 1              ~&~ for $\kp\ll1$; \cr
               (2\kp)^{-1} ~&~ for $1<2\kp<m_p/m_e$;\cr
               m_e/m_p              ~&~ for $2\kp>m_p/m_e$. \cr}
\label{eq:mu}
\end{eqnarray}
Provided that the characteristic Lorentz factor of leptons is not
less than that of baryons, the $\xi_e$ value is limited to the
regime
\begin{eqnarray}
\begin{array}{cl}
2\kp m_e/m_p\la\xi_e\la1 & {\rm ~for~} 2\kp<m_p/m_e,\\
\xi_e\approx1 & {\rm ~for~} 2\kp>m_p/m_e.
\end{array}
\end{eqnarray}

%%%%%%%%%%%%%%%%%%%%%%%%%%%%%%%%%%%%%%%%%%%%%%%%%%%%%%%%%%%%%%%%%%%%%%%%%%%%%%%%%%%%%%%%%%%%%%%%
\subsection{The $1<2\kp<m_p/m_e$ case}
If $\kp\gg1$, the random Lorentz factor of leptons becomes lower
since the lepton number increases significantly. However, the
pairs do not change the magnetic field in region 3, which is
assumed to have an energy density $\xi_B$ times of the internal
energy density, that is, $B=\gamma_3c(32\pi n_1m_p\xi_B)^{1/2}$
with $n_1$ being the medium number density. In low-$\sigma$ case,
the magnetic field may be shock-induced, thus we assume
$\xi_B\sim10^{-2}$, as is apparent in the long-term forward
shock-produced afterglow. But in high-$\sigma$ case it may be
dominated by the original wind magnetic field, we then expect much
higher $\xi_B$ value. Thus, for $1<2\kp<m_p/m_e$, the observed
peak frequency of synchrotron photons $\nu_m\propto\gamma_m^2B$ is
lower by a factor of $(2\kp)^{-2}$. Substituting Eqs. (\ref{gm})
and (\ref{eq:mu}) we have
\begin{eqnarray}\label{num}
\nu_m&=&\gamma_3\gamma_m^2\frac{eB}{2\pi m_ec}\nonumber\\
%&=&3.6\times10^{10}\xi_{e,-1}^2\xi_{B}^{1/2}n_1^{1/2}\k2^{-2}\gamma_{3,300}^2{\rm Hz},
&=&8.2\times10^{10}\xi_{e,-1}^2\xi_{B}^{1/2}n_1^{1/2}\k2^{-2}\gamma_{4,300}^2{\rm
Hz},
\end{eqnarray}
where $\gamma_3$ is cancelled out in the caculation. Note that the
emission is in quite low frequencies, in contrast with the
optical/UV for the case without pairs $(\mu m_e/m_p\approx 1)$.
The cooling frequency is relevant to those leptons cool at the
dynamical time $t$. For $t=t_{cr}$ it is
\begin{equation}\label{nuc}
\nu_c=1.1\times10^{14}\xi_B^{-3/2}E_{a,52}^{-1/2}n_1^{-1}t_{cr,1}^{-1/2}{\rm
Hz},
\end{equation}
most sensitive to $\xi_B$. If $\xi_B\sim10^{-2}$, $\nu_c$ is
larger, $\sim10^{17}$~Hz.

The pairs also significantly increase the radiative efficiency by
increasing the radiating lepton number. At the crossing time
$t_{cr}$, the reverse shock has swept up all the shell leptons,
and the observed emission reaches a peak. The total lepton number
is $N_e\equiv N_b(1+2\kp)\simeq2N_\pm$, and, with help of Eq.
(\ref{gamma3}), the peak flux is
\begin{eqnarray}\label{fluxm}
F_{\nu_m}&= & \gamma_3N_e \frac {e^3B}{4\pi d_l^2m_ec^2}\nonumber\\
&\simeq &
%1.7\times10^3~\xi_{B}^{1/2}n_1^{1/2}d_{l,28}^{-2}E_{a,52}\k2\gamma_{3,300}{\rm ~Jy},
960~\frac{ \xi_{B}^{1/2}n_1^{1/4}E_{52}E_{a,52}^{1/4}\k2}
{d_{l,28}^{2}\Gamma_{300}(1+\sigma)t_{cr,1}^{3/4}} {\rm ~Jy},
\end{eqnarray}
where $d_l$ is the GRB luminosity distance. The flux increases by
a factor of $2\kp$ as opposed to the case without pair effect.

The synchrotron spectrum of power-law-distributed leptons is
simply described by power-law segments: (1) For $\nu<\nu_m$ we
have the synchrotron low-energy tail $F_\nu\propto\nu^{1/3}$. (2)
For $\nu_m<\nu<\nu_c$ we have the typical synchrotron slope
depending on the lepton's distribution,
$F_\nu\propto\nu^{-(p-1)/2}$. (3) For $\nu>\nu_c$ the spectrum
behaves as a fast-cooling slope, $F_\nu\propto\nu^{-p/2}$. We are
particularly interested in the optical prompt emission, say, in
the R band at $\nu_R=4.5\times10^{14}$~Hz, a frequency well above
the characteristic frequency $\nu_m$ and around the cooling
frequency $\nu_c$. Because of the independence of $\nu_c$ on
$\kp$, no matter whether $\nu_R$ is above $\nu_c$ or not we
usually have $F_{\nu_R}\propto
F_{\nu_m}\nu_m^{(p-1)/2}\propto\kp^{-(p-2)}$. Since $p\approx2$,
this means that the emission is insensitive to pair loading. For
simplicity $p=2$, we have, for small $\xi_B$ or $\nu_R<\nu_c$,
\begin{equation}\label{fluxnuR<nuc}
%15\xi_{e,-1}\xi_B^{3/4}n_1^{3/4}d_{l,28}^{-2}E_{a,52}\gamma_{3,300}^2{\rm~Jy},&{\rm if~} \nu_R<\nu_c,\\
F_{\nu_R<\nu_c}=0.41\frac{\xi_{e,-1}\xi_{B,-2}^{3/4}n_1^{1/2}E_{52}E_{a,52}^{1/4}\gamma_{4,300}}
{d_{l,28}^{2}\Gamma_{300}(1+\sigma)t_{cr,1}^{3/4}} {\rm ~Jy},
\end{equation}
and for enough large $\xi_B$ or $\nu_R>\nu_c$,
\begin{equation}\label{fluxnuR>nuc}
F_{\nu_R>\nu_c}=
5.8\frac{\xi_{e,-1}E_{52}\gamma_{4,300}}{d_{l,28}^{2}\Gamma_{300}(1+\sigma)t_{cr,1}}{\rm~Jy},
\end{equation}
where $\kp$ is cancelled out, and $\Gamma$ and $\gamma_4$ can be
also cancelled out if $\Gamma\sim\gamma_4$.

Synchrotron self-absorption may be important at low frequencies
here. We follow the simple way by Sari \& Piran (1999b) and
Chevalier \& Li (2000) to estimate the maximal flux emitted by the
shocked shell material as a blackbody
\begin{equation}\label{fluxbb}
F_{\nu,bb}\approx\pi\pfrac{R_\perp}{d_l}^2\frac{2\nu^2}{c^2}kT_{eff},
\end{equation}
with the observed size of the fireball $R_\perp$ given roughly by
\begin{equation}\label{Rsize}
R_\perp\approx2\gamma_3ct
\end{equation}
and the effective temperature by
\begin{equation}\label{teff}
kT_{eff}\approx\gamma_3\gamma_\nu m_ec^2/3,
\end{equation}
where $\gamma_\nu$ is the Lorentz factor of the electrons that
radiate at the frequency $\nu$ and is given by $(2\pi
m_ec\nu/\gamma_3 eB)^{1/2}$. Substituting Eqs. (\ref{gamma3}),
(\ref{Rsize}) and (\ref{teff}) into Eq. (\ref{fluxbb}) leads to
\begin{equation}\label{<fluxbb}
F_\nu\leq
%F_{\nu,bb}\approx2.5\pfrac\nu{\nu_R}^{5/2}\xi_B^{-1/4}n_1^{-1/4}d_{l,28}^{-2}\gamma_{3,300}^2t_{cr}^2{\rm Jy},
F_{\nu,bb}\approx150\pfrac\nu{\nu_R}^{5/2}\xi_B^{-1/4}E_{a,52}^{1/4}n_1^{-1/2}d_{l,28}^{-2}t_{cr,1}^{5/4}{\rm
Jy},
\end{equation}
where we scale the observed frequency in units of R band frequency
$\nu_R$. It is obvious that the peak flux in characteristic
frequency $\nu_m$ usually suffers strong self-absorption. If the
absorption frequency $\nu_a$ is defined by $F_\nu=F_{\nu,bb}$, we
should have $\nu_a>\nu_m$. Assuming further $\nu_a<\nu_c$, we can
calculate its value,
\begin{equation}
\nu_a\approx2.0\times10^{14}\pfrac{\xi_{e,-1}\xi_Bn_1E_{52}\gamma_{4,300}}{\Gamma_{300}(1+\sigma)t_{cr,1}^2}^{1/3}{\rm
Hz}.
\end{equation}
This is in the IR band, $\nu_a\sim4\times10^{13}$~Hz, for
$\xi_B\sim10^{-2}$. As shown in Fig. \ref{fig:spectrum}, the
observed spectrum peaks at $\nu_a$, and below $\nu_a$ it behaves
as $F_\nu=F_{\nu,bb}\propto\nu^{5/2}$.

%%%%%%%%%%%%%%%%%%%%%%%%%%%%%%%%%%%%%%%%%%%%%%%%%%%%%%%%%%%%%%%%%%%%%%%%%%%%%%%%%%%
\subsection{The $2\kp>m_p/m_e$ case with high $\sigma$}
A low-$\sigma$ GRB usually do not produce \pair~pairs with mass
larger than fireball baryons, $2\kp>m_p/m_e\sim 1800$, because
that $R_{dom}<R_b$ (see Fig. \ref{fig:radius}). However, if
$\sigma\ga1$ where the initial wind/fireball is dominated by
magnetic field, it may appear that $R_{dom}>R_b$. So, the
$2\kp>m_p/m_e$ case may emerge below $R_{dom}$ and above $R_b$ in
Fig. \ref{fig:radius} for high $\sigma$. It is usually apparent
that $R_{dom}<R_\pm$, except for $\sigma\ga10^2$, therefore the
case $2\kp>m_p/m_e$ should be associated with optically thick pair
screens, and, as suggested by \Mesz \etal (2002), accompanied with
X-ray flashes\footnote{The generated pair number is prevented from
multiplying beyond a value corresponding to $\tau_\pm$ of a few,
due to self-shielding. If $\tau_\pm<10$ is required, the regime
with $R<R_\pm/\sqrt{10}$ is not relevant to any observed bursts.
So the $2\kp>m_p/m_e$ case, which requires $R<R_{dom}$, do not
emerge unless $R_{dom}>R_\pm/\sqrt{10}$ for high $\sigma$.}.

%From Eq. (\ref{number}) we can see that the condition
%$2\kp>m_p/m_e\sim 1800$ is easy to achieve, provided somewhat
%larger energy dissipation $E_\gamma$ or somewhat smaller Lorentz
%factor $\Gamma$ and emission radius $R$. Meanwhile, these may lead
%to much smaller $R_{var}$ (Eq. (\ref{Rvar})) and larger $R_\pm$
%(Eq. (\ref{Rpair})), and the optical-thick case of $R\leq
%R_{var}\la R_\pm$ may appear. \Mesz \etal (2002) have argued that
%for emission sites of $R<R_\pm$, the photons will suffer
%multi-Compton scattering before escaping, leading to current known
%X-ray flashes. Therefore, though not strictly speaking, we expect
%that a $2\kp>m_p/m_e$ case and an X-ray flash are likely to emerge
%together. For example, for smaller Lorentz factor $\Gamma\sim60$
%and then smaller emission site $R\la R_{var}\sim2\times10^{13}$~cm
%(Eq. (\ref{Rvar})) , we have $\kp\sim10^4$ (Eq. (\ref{number}))
%and $R_{var}<R_\pm\sim4\times10^{14}$~cm (Eq. (\ref{Rpair})).

For $\xi_e\approx1$ and $\mu\approx1$ in this case, from Eq.
(\ref{gm}) the characteristic lepton Lorentz factor is simply
$\gamma_m\approx\gamma_4/\gamma_3$, which is independent of the
pair number or $\kp$. The corresponding emitted frequency is
\begin{equation}
\nu_m\simeq1.1\times10^{10}\xi_B^{1/2}n_1^{1/2}\gamma_{4,100}^2{\rm
~Hz},
\end{equation}
independent of $\xi_e$ and $\kp$, where we have scaled
$\gamma_{4,100}=\gamma_4/100$ for regime below $R_{dom}$ in Fig.
\ref{fig:radius}. At the same time, the Eqs. (\ref{nuc}),
(\ref{fluxm}) and (\ref{<fluxbb}) above are still valid. It should
be noticed that the peak time $t_{cr}$ is larger for lower Lorentz
factor, $t_{cr}\geq
t_{dec}=63E_{a,52}^{1/3}n_1^{-1/3}\gamma_{4,100}^{-8/3}$~s. We
calculate the R-band emission, which is dependent on $\kp$,
$F_{\nu_R}\propto F_{\nu_m}\nu_m^{(p-1)/2}\propto\kp$. It is, for
$\nu_R<\nu_c$,
\begin{equation}
F_{\nu_R<\nu_c}=0.80\frac{\xi_{B,-2}^{3/4}n_1^{1/2}E_{52}E_{a,52}^{1/4}\gamma_{4,100}{\kp}_4}
{d_{l,28}^{2}\Gamma_{100}\sigma_1t_{cr,2}^{3/4}} {\rm ~Jy},
\end{equation}
or for $\nu_R>\nu_c$,
\begin{equation}
F_{\nu_R>\nu_c}=
6.9\frac{E_{52}\gamma_{4,100}{\kp}_4}{d_{l,28}^{2}\Gamma_{100}\sigma_1t_{cr,2}}{\rm~Jy},
\end{equation}
where we have taken $1+\sigma\simeq\sigma=10\sigma_1$. Note that
$\kp$ and $t_{cr}$ are not free and are determined by other
parameters.

%%%%%%%%%%%%%%%%%%%%%%%%%%%%%%%%%%%%%%%%%%%%%%%%%%%%%%%%%%%%%%%%%%%%%%%%%
\subsection{Comparision of spectral profiles with $\kp\ll1$ case}
Regardless of pair effects, the typical frequency of the reverse
shock falls in the optical/UV regime, and self-absorption can
hardly play a role at the optical (e.g. Sari \& Piran 1999b). So
the flux behaves as the synchrotron tail $F_\nu\propto\nu^{1/3}$
just below the optical, and $F_\nu\propto\nu^2$ at somewhat lower
frequencies where self-absorption is important.

For pari-rich reverse shocks, the typical frequency is far below
the optical and the emission there suffer strong self-absorption,
so a steep slope 5/2 is apparent in the self-absorption regime of
lower frequencies. The spectral indices from low to high
frequencies are in the order of $[5/2,\sim-1/2,\sim-1]$ for
$\nu_a<\nu_c$ (or $[5/2,\sim-1]$ if $\nu_a>\nu_c$), as shown in
Fig. \ref{fig:spectrum}. We note that the spectral index 5/2 and
stronger fluxes below the optical are the distinct properties of a
pair-rich reverse shock. Future spectroscopy observation of prompt
emission in longer wavelengths, such as by {\em REM} (Zerbi \etal
2001), may detect the 5/2 slope in IR spectra, and reveal the pair
influence.

%%%%%%%%%%%%%%%%%%%%%%%%%%%%%%%%%%%%%%%%%%%%%%%%%%%%%%%%%%%%%%%%%%%%%%%%%%
\section{Observations}

ROTSE detected a strong optical flash during GRB 990123, which
reached a peak of $\sim 1$~Jy 50 s after the GRB trigger (Akerlof
\etal 1999). From Eqs. (\ref{fluxnuR<nuc}) and (\ref{fluxnuR>nuc})
a pair-rich reverse shock is easy to emit such a strong flash.
Absorption lines in the optical afterglow have inferred a redshift
of $z=1.6$, or a luminosity distance of $d_l=4\times10^{28}$~cm
(in a $H_0=71$~km s$^{-1}$Mpc$^{-1}$, $\Omega_m=0.27$ and
$\Omega_\Lambda=0.73$ universe). The released gamma-ray energy is
therefore huge, $E_\gamma\sim10^{54}$~ergs, so we assume $E\sim
E_a\sim10^{54}$~ergs. Noting that the peak time is
$t_{cr}\sim50/(1+z)\sim20$~s, the other reasonable parameters to
achieve a 1-Jy flash are in their typical values:
$\xi_{e,-1}\xi_{B,-2}^{3/4}n_1^{1/2}\gamma_{4,300}
\Gamma_{300}^{-1}(1+\sigma)^{-1}\approx1$.

However, strong optical flashes appear to be rare, since they have
not been detected from other bursts (Akerlof \etal 2000; Kehoe
\etal 2001). Especially, GRBs 981121 and 981223 are observed to
have flashes, if any, fainter than $\sim10$~mJy a few tens seconds
after the GRB triggers. To solve this problem in the pair-poor
reverse shock model, one needs to assume that the characteristic
frequency is either well above or below the optical,
$\nu_m\gg\nu_R$ or $\nu_m\ll\nu_R$ (Kobayashi 2000). However
Soderberg \& Ramirez-Ruiz (2002) have investigated a subset of
GRBs with afterglow-determined parameters, and shown that strong
flashes may be expected in all of these bursts.

For pair-rich reverse shocks, it is usually the case of strong
flashes in which the optical emission is independent of the pair
numbers (Eqs. (\ref{fluxnuR<nuc}) and (\ref{fluxnuR>nuc})). Since
the gamma-ray fluences of GRBs 981121 and 981223 is about one or
two orders of magnitude lower than GRB 990123, their released
gamma-ray energy is $E_\gamma\sim10^{52}$~ergs (Kobayashi 2000) if
assuming comparable redshift $z\sim2$ ($d_{l,28}\sim5$). To
achieve a faint flux, $<10$~mJy, in Eq. (\ref{fluxnuR<nuc}) one
needs some extreme parameter values, e.g. $\xi_B<10^{-3}$.  But we
expect the fluxes are generally large for typical parameters. The
rarety of strong flashes in observation may need some external
reasons to account for, for example, dust obscuration of the
optical/UV in the star formation regions.

%%%%%%%%%%%%%%%%%%%%%%%%%%%%%%%%%%%%%%%%%%%%%%%%%%%%%%%%%%%%
\section{Conclusions and discussion}

In this work we calculate the \pair ~pair number produced in GRB
fireballs, and study their influence on the reverse-shock
emission. We have assumed that the intrinsic GRB spectrum, before
$\gamma-\gamma$ absorption, can be extrapolated to very high
energy, $\eps_{max}\gg\eps_{cut}$. Our calculation of \pair~pair
number does not depend on any special model, and is only based on
observations. We show that, for typical parameters, the GRB
wind/fireball will become pair-rich during the bursting phase,
owing to intense \pair~pair production. Later on, the pairs are
carried into the reverse shock and give rise to the intense prompt
flash which peaks at lower frequency, i.e. in the IR band, as
opposed to the optical/UV emission in the case where pair-loading
is negligible. The emission below the observed peak frequency
suffers strong self-absorption, leading to the distinct
self-absorption spectral index 5/2.
%{\bf Since the pairs are produced in GRBs and radiating in  flashes, the pairs bring a link between GRBs and early
%afterglows.}

\begin{figure}
\centerline{\hbox{\psfig{figure=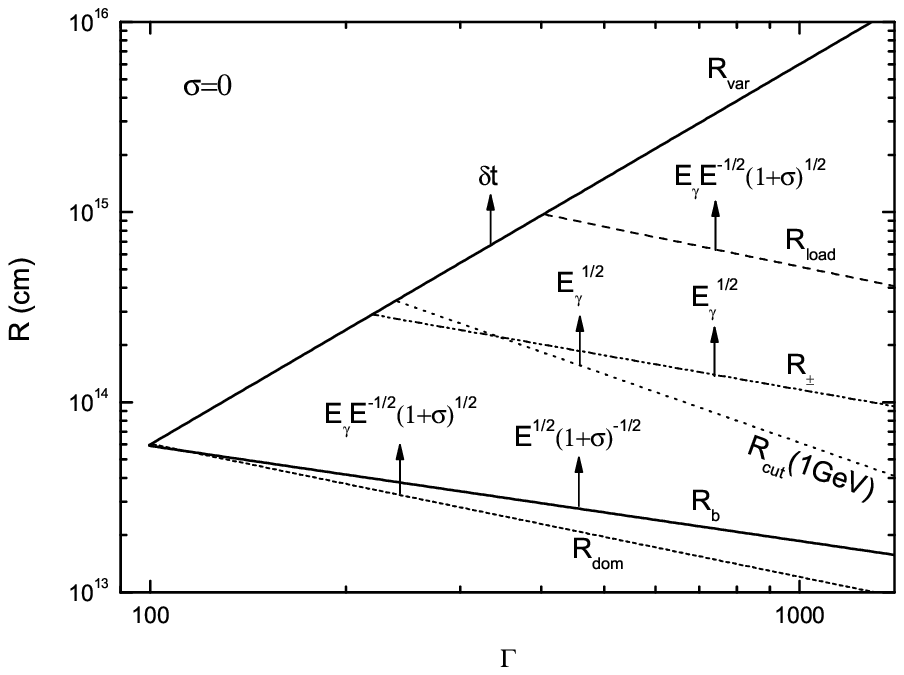,width=4.8in,angle=0}}}
\centerline{\hbox{\psfig{figure=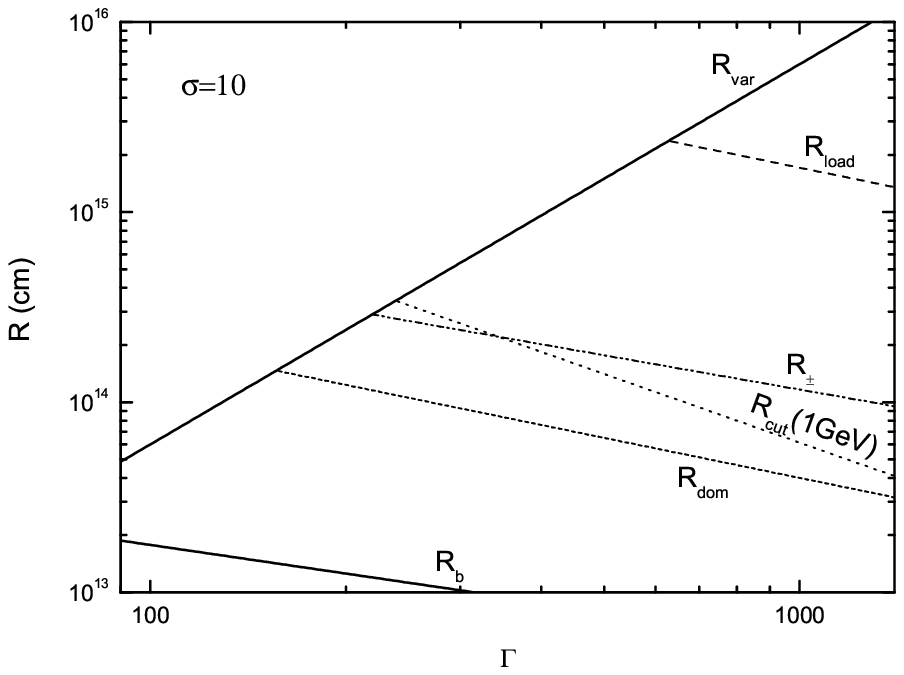,width=4.8in,angle=0}}}
\caption{Emission site $R$ versus initial Lorentz factor $\Gamma$:
$\sigma=0$ ({\it upper frame}) and $\sigma=10$ ({\it lower
frame}). Neither the regime above $R_{var}$ nor that below $R_b$
is relevant to observed GRBs. Below $R_{load}$ is the regime where
pair loading is important, $2\kp>1$. Case $2\kp>m_p/m_e$ is even
available below $R_{dom}$. A classic GRB, with non-thermal sub-MeV
radiation and rapid variabilities, should be optically thin to
scattering of photons by baryonic electrons ($R>R_b$) and
secondary electron-positron pairs ($R>R_\pm$). The X-ray flashes
are possible to appear below $R_\pm$. The GRBs shows no
attenuation above 1 GeV correspond to regime beyond the curve
marked $R_{cut}$(1 GeV). $E_{52}=E_{\gamma,52}=\eps_b=\delta
t_{-1}=1$ and $\beta=2.2$ are assumed. The scaling laws of special
radii with the parameters are also marked, except for $\eps_b$ to
which the relation is usually weak.}\label{fig:radius}
\end{figure}

\begin{figure}
\centerline{\hbox{\psfig{figure=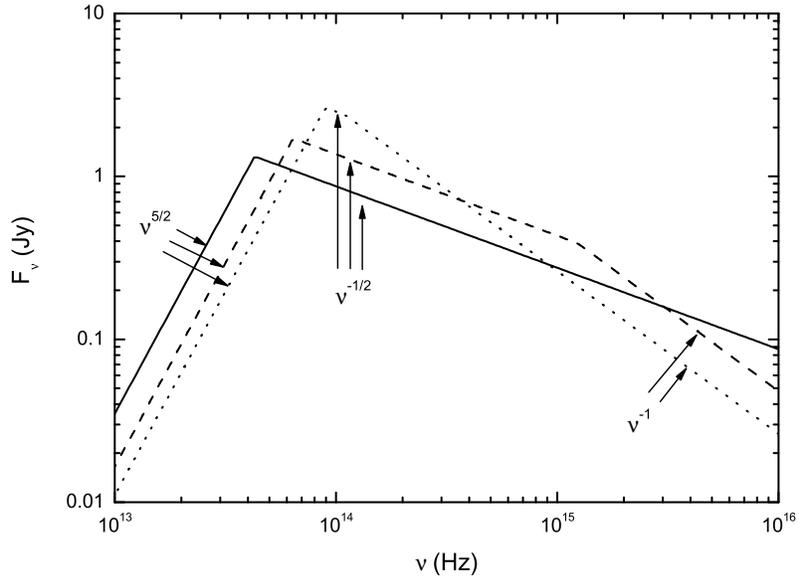,width=4.8in,angle=0}}}
\caption{Examples of prompt-emission spectra from pair-rich
reverse shocks: $(\xi_B, \sigma)=(0.01, 0)$ ({\it solid line}),
$(0.2, 5)$ ({\it dashed line}) and $(1, 10)$ ({\it dotted line}),
respectively. Other typical parameter values are assumed (see the
text), especially that $\k2=t_{cr,1}=1$.}\label{fig:spectrum}
\end{figure}

Since $\kp\propto\Gamma^{-(2\beta-3)}R^{-2}$, pair loading is
unimportant for Lorentz factors large enough (also see Fig.
\ref{fig:radius}). However, the Lorentz factors of GRBs have been
found not to be quite large by some previous works. For examples,
in the internal shock model of GRBs, the burst emission should
come from internal shocks before significant deceleration due to
the swept-up external matter, which results in $\Gamma\lesssim
10^3$ (Lazzaiti, Ghisellini \& Celotti 1999); The observed pulse
width evolution in the bursting phase seems to rule out the GRB
models with $\Gamma\gg10^3$ (Ramirez-Ruiz \& Fenimore 2000); For
GRBs to be efficient, the cooling timescale of leptons should be
smaller than the GRB duration, leading to an upper limit of
$\sim1200$ (Derishev, Kocharovsky \& Kocharovsky 2001). We,
therefore, expect that most GRBs are pair-rich with $2\kp>1$.

The pair-rich reverse shocks emit mainly in the IR band. If GRBs
are associated with star forming regions, as implied by the
observations (e.g., Bloom \etal 1999; Fruchter \etal 1999; Bloom,
Kulkarni \& Djorgovski 2001) and by the favored progenitor models
that GRBs come from explosions of massive stars (Woosley 1993;
\Pacz 1998; Vietri \& Stella 1999; MacFadyen \& Woosley 1999), the
optical may suffer dust obscuration, whereas the IR would not
(e.g., Draine \& Hao 2002; Reichart \& Yost 2002; Ramirez-Ruiz
\etal 2002; Fruchter \etal 2001; Waxman \& Draine 2000). Strong
pair-loading also increase significantly the flash brightness.
Thus, we expect that there are many strong IR flashes associated
with GRBs (and XRFs). If so, the rarity of optical flashes in GRBs
will infer dust-enshrouded environment of GRBs.  The detection of
prompt IR flashes requires rapid response to the GRB trigger in
such band, and the up-coming near-IR telescope, {\it REM} (Zerbi
\etal 2001), is capable of catching them.

The observation in high energy range, e.g., $\gtrsim100$~MeV,
plays an important role in GRB research, because they provide a
way to determine the initial Lorentz factor $\Gamma$ (e.g. Baring
2000; Lithwick \& Sari 2001, and references therein). If a GRB
exhibits attenuation above some $\eps_{cut}$, we can use
Eq.(\ref{cut}) together with Eq. (\ref{Rvar}) to constrain
$\Gamma$. If $\eps_{cut}\approx1$~GeV, then we have, for typical
parameters, $\Gamma\ga230$ where the equality is for $R\approx
R_{var}$ (see the crossing point between curves $R_{cut}$ and
$R_{var}$ in Fig. \ref{fig:radius}). The high-resolution spectra
of GRBs may be provided by the next generation high-energy
gamma-ray observatory, e.g., {\em GLAST}.

Eq.(\ref{number1}) indicates that $\sim10^{54}E_{\gamma,52}^2$
pairs are produced in a GRB. This number should be reduced by a
factor of $\theta_{jet}^2/4$ if GRBs are beamed. It is recently
suggested that the annihilation of the relic positrons provides
identification beacon for at least one GRB remnant in the Milky
way (Furlanetto \& Loeb 2002).

%%%%%%%%%%%%%%%%%%%%%%%%%%%%%%%%%%%%%%%%%%%%%%%%%%%%%%%%%%%%
\acknowledgments

We would like to thank the referee for valuable
comments. Z. Li thanks D.M. Wei, X.Y. Wang, Y.F. Huang and X.F. Wu
for helpful discussions. This work was supported by the National
Natural Science Foundation of China (grants 19825109 and
19973003), the National 973 Project (NKBRSF G19990754) and the
Special Funds for Major State Basic Research Projects.

\end{document}